\documentclass[pdftex,twocolumn,epjc3]{svjour3}          

\RequirePackage[T1]{fontenc}
\RequirePackage{graphicx}
\RequirePackage{subfigure}
\RequirePackage{epstopdf}
\RequirePackage{tabularx}
\RequirePackage{amssymb}
\RequirePackage{amsmath}
\usepackage{color}

\RequirePackage{feynmp}
 \RequirePackage[utf8]{inputenc}

\RequirePackage{float}
\RequirePackage{wasysym}

\journalname{Eur. Phys. J. C}

\newcommand{\yu}{y_U}
\newcommand{\yd}{y_D}
\newcommand{\ye}{y_E}
\newcommand{\yt}{y_t}
\newcommand{\yb}{y_b}
\newcommand{\ytau}{y_\tau}
 \newcommand{\lu}{\lambda_t}

\newcommand{\lsd}{\lambda_{S_d}}
\newcommand{\GeV}{\, {\rm GeV}}

\begin{document}

\title{Extended Gauge Mediation in the NMSSM with Displaced LHC Signals} 

\author{
Marcin Badziak\thanksref{a, addr1} \and
Nishita Desai\thanksref{b, addr2} \and
Cyril Hugonie\thanksref{c, addr2} \and
Robert Ziegler\thanksref{d, addr3}}
\thankstext{a}{mbadziak@fuw.edu.pl}
\thankstext{b}{nishita.desai@umontpellier.fr}
\thankstext{c}{cyril.hugonie@umontpellier.fr}
\thankstext{d}{robert.ziegler@cern.ch}

\institute{
\label{addr1} Institute of Theoretical Physics, Faculty of Physics,
  University of Warsaw, ul.\ Pasteura 5, PL-02-093 Warsaw, Poland \and
\label{addr2} LUPM, UMR 5299, CNRS, Universit\'{e} de Montpellier, 34095, Montpellier, France \and
\label{addr3} Theoretical Physics Department, CERN, 1 Esplanade des Particules, Geneva 23, CH-1211, Switzerland
}

\maketitle


\begin{abstract}
We analyze models of extended Gauge Mediation in the context of the NMSSM, concentrating on supersymmetric spectra with light gluinos, low fine-tuning and decays of the lightest neutralino leading to displaced vertices. While the minimal scenario has rather heavy gluinos as a result of restrictions from the Higgs sector, we propose two new models in which the gluino can be as light as allowed by direct searches at the LHC, with a mass of about  $1.7 \, {\rm TeV}$ and  $2.0 \, {\rm TeV}$, respectively. Both models have a tuning of a few permille, and lead to an interesting phenomenology due to a light singlet sector. A singlet state at around 98 GeV can account for the LEP excess, while the singlino has a mass of the order of 100 GeV and decays to b-jets and the  gravitino, with decay lengths of a few cm.  
\end{abstract}

\section{Introduction}

Gauge mediated supersymmetry breaking (GMSB) is one of the most elegant ways to explain the absence of sizable contributions to flavour violating processes in supersymmetric (SUSY) models. However, its simplest realizations are inconsistent with the measured Higgs mass of 125 GeV, unless one is willing to accept a very heavy (and thus unnatural) SUSY spectrum. In general there are two ways to make GMSB viable with sparticles light enough to be in the LHC reach. The first option to boost the Higgs mass is to maximize the loop corrections with sizable stop mixing, which requires direct Higgs-messenger couplings \cite{Evans:2011bea,Jelinski:2011xe,Evans:2012hg,Kang:2012ra,Craig:2012xp,Abdullah:2012tq,Kim:2012vz,Byakti:2013ti,Craig:2013wga,Evans:2013kxa,Calibbi:2013mka,Jelinski:2013kta,Galon:2013jba,Fischler:2013tva,Knapen:2013zla,Ding:2013pya,Calibbi:2014yha,Basirnia:2015vga,Jelinski:2015gsa,Jelinski:2015voa,Knapen:2015qba}. The second (less explored) option is to introduce additional tree-level contributions 
to the Higgs mass. Such corrections  may originate from the mixing of the Higgs with a lighter singlet~\cite{Badziak:2013bda}, and can be realized in a simple model proposed by Delgado, Giudice and Slavich  (DGS)~\cite{Delgado:2007rz} that combines the NMSSM and GMSB with direct singlet-messenger couplings. Indeed it  has been shown in Ref.~\cite{Allanach:2015cia} that the most interesting region in parameter space features a light SUSY spectrum and a singlet around 90 $\div$100 GeV, which gives  a sizable tree-level contribution to the SM-like Higgs mass through mixing. Besides this ``push-up" of  the Higgs mass, the general framework of NMSSM+GMSB  with a light singlet has other advantages. First, it provides a natural solution to the $\mu$-$B_\mu$ problem~\cite{Delgado:2007rz}, as $\mu$ and $B_\mu$  are generated dynamically through the vacuum expectation value (VEV) of the singlet. Second, the Next-to-Lightest SUSY particle (NLSP) 
is typically the singlino, whose decay into the gravitino is suppressed, leading to  novel displaced signatures at colliders~\cite{Allanach:2016pam}.

In the present paper, we focus on GMSB + NMSSM models with direct matter-messenger couplings, with a special emphasis on displaced signatures. This is motivated by recent LHC results for heavy Higgs searches in the $\tau\tau$ channel, which has excluded the most interesting part of the parameter space of the DGS model. We investigate two simple extensions of the DGS model leading to a relatively light sparticle spectrum, which can be probed in the second run of the LHC, in particular associated  with displaced decay signatures.

The rest of the article is organized as follows: In section~\ref{sec:EWSB} we analyse the general features of SUSY models with light singlets. In section~\ref{sec:DGS} we review the DGS model and investigate the impact of recent LHC results on its parameter space. In  sections~\ref{Umodel} and \ref{DGSUmodel}  we introduce two new models with singlet-messenger couplings, and summarize and conclude in section~\ref{conclusions}. In an Appendix we provide complete expressions for soft terms in the respective models. 

\section{EWSB in the NMSSM with a Light Singlet}
\label{sec:EWSB}

Let us start with a brief discussion of Electroweak Symmetry Breaking (EWSB) in the $\mathbb{Z}_3$ invariant NMSSM, in which the $\mu$-term in the MSSM superpotential is replaced by the following singlet couplings
\begin{equation}
\label{W_NMSSM}
 W_{\rm NMSSM}= \lambda SH_uH_d + \frac13\kappa S^3 \,,
\end{equation} 
and the NMSSM specific soft terms are given by 
\begin{align}
 -\mathcal{L}_{\rm soft} & \supset  m_{H_u}^2 |H_u|^2 + m_{H_d}^2 |H_d|^2 + m_S^2 |S|^2 +  \nonumber \\ 
  & + ( A_{\lambda} \lambda H_u H_d S +\frac{1}{3}\kappa A_{\kappa}
S^3  + {\rm h.c.} )\,. 
\end{align}
In order to generate a sufficiently large VEV for the singlet $\langle S \rangle\equiv s$,
the singlet soft mass $m_S^2$ must be negative or at least much smaller than $A_\kappa^2$. This statement can be quantified in the limit
$s\gg v$, where $v$ is the electroweak scale, in which one finds
\begin{align}
\label{eq:def_wz}
s & \approx \frac{A_\kappa}{\kappa} w \, , & w & \equiv \frac{1 + \sqrt{1- 8z} }{4} \, , & z & \equiv \frac{m_S^2}{A_\kappa^2} \,.
\end{align}
The approximate condition for proper EWSB in a global minimum of the potential reads
\begin{equation}
\label{eq:Svev_cond}
  z \lesssim \frac19\ \Leftrightarrow \  w \gtrsim \frac13\,.
\end{equation}
In Minimal Gauge Mediation soft terms for the singlet sector do not arise at leading order which prevents successful EWSB.
However, sufficiently large soft terms can be generated if there are direct couplings of the singlet to the messenger sector, as originally proposed in Ref.~\cite{Giudice:1997ni} and worked out in detail in Ref.~\cite{Delgado:2007rz}. It was further demonstrated in Ref.~\cite{Allanach:2015cia} that this model allows to realize the interesting ``push-up" scenario where the SM-like Higgs mass gets a large positive contribution from mixing with a {\it lighter} singlet-like Higgs, thus allowing to lower the overall scale of  the SUSY spectrum that drives the radiative corrections to the Higgs mass.

In the following we revisit this model (DGS) in the light of updated experimental constraints, and compare it to the phenomenology of two similar models that feature additional singlet-messenger couplings. We are particularly
interested in SUSY spectra with experimental signatures that may be probed at the LHC using displaced vertices, as studied in Ref.~\cite{Allanach:2016pam}. 

Before discussing these models, we note that in their most interesting regions of the parameter space, where the SUSY spectrum can be relatively light thanks to the push-up effect of Higgs-singlet mixing,  correct EWSB generically implies large values of $\tan\beta$. As a consequence, the most stringent constraints on these models often come from LHC searches for
heavy MSSM-like Higgs bosons $H/A$ decaying to $\tau\tau$, because the cross-section for $H/A$ production grows with $ \tan^2 \beta$.
Since this constraint turns out so important, it is instructive to discuss the origin of large $\tan \beta$ in the push-up region of NMSSM models with Gauge Mediation. Neglecting
$m_{H_d}^2$, which is usually a good approximation in the models under consideration, the expression for $\tan\beta$ is relatively simple:
\begin{align}
\tan\beta & \approx \frac{\lambda}{\kappa} \frac{A_\kappa w}{A_\kappa w - A_\lambda} \, , & \frac{\lambda^2}{\kappa^2} & \approx \frac{(A_\kappa w - A_\lambda )^2 -   m_{H_u}^2}{A_\kappa^2 w^2}   \, .
\end{align}
The soft parameters can in turn be related to the physical Higgs boson masses using the following approximate expressions for CP-odd Higgs masses:
\begin{align}
m_{a1}^2 & \approx 3 A_\kappa^2 w \, , &  m_{a2}^2 & \approx m_{H_d}^2-m_{H_u}^2  \approx -m_{H_u}^2 \, , 
\end{align}
and the lightest (singlet-like) CP-even Higgs mass:
\begin{align}
m_{h_1}^2 = A_\kappa^2 w (4 w -1) \, .
\end{align}
We also note that for $\mu>0$, which we always assume in this analysis\footnote{Also $\mu< 0$ leads to viable spectra, but we will not consider these scenarios in the following, since the gluino is always quite heavy with a mass above 3.2 TeV.}, $A_\kappa$ must be negative to avoid tachyons in the singlet sector.
In the limit $w\gg1$, corresponding to the case where EWSB is driven by a large negative $m_S^2$, one finds indeed large values of $\tan \beta$
\begin{equation}
 \label{eq:tanbeta_largew}
\tan\beta \approx \frac{\lambda}{\kappa} \approx \frac{|m_{H_u}|}{|A_\kappa| w} \approx  \frac{2m_{a2}}{m_{h1}} \gg 1 \, .
\end{equation}
Using this relationship between $\tan \beta$ and $m_{a2}$  at face value with $m_{h1}\approx100$~GeV, the recent ATLAS constraint from Ref.~\cite{Aaboud:2017sjh}, implies a limit $\tan\beta\lesssim25$ or equivalently
$m_{a2}\gtrsim1.2$~TeV, assuming no heavy Higgs decays to SUSY particles and neglecting threshold corrections to bottom quark Yukawa
couplings. In typical GMSB models the value of $m_{a2}$ is correlated to other sparticle masses,  including squarks and gluino, so that a stronger
bound on $m_{a2}$ typically results in a stronger bound on coloured sparticles. 

For smaller $w$ the value of $\tan\beta$ slightly decreases. In the limit $w\approx1/3$ one finds
\begin{equation}
\label{eq:tanbeta_w13}
\tan\beta \approx \frac{|m_{a2}|}{m_{h1}} \frac{1}{1-  3 A_\lambda/A_\kappa} \approx \frac{|m_{a2}|}{m_{h1}} \frac{1}{1+  A_\lambda/m_{h1}} \gg 1 \, ,
\end{equation}
Neglecting the terms proportional to $A_\lambda$, $\tan\beta$ is smaller by a factor two with respect to the limit $w\gg1$, so that the LHC constraints on the $a_2$ are
expected to be relaxed. Taking into account non-zero $A_\lambda$, one can suppress (enhance) $\tan\beta$ when  $A_\lambda$ is positive (negative).
We also note that $A_\lambda$ may get positive contributions via RG running from a negative top trilinear term $A_t$.

Finally we provide an approximate expression for the $Z$-boson mass that is convenient to assess the fine-tuning. In the large $\tan \beta$ limit one has
\begin{align}
M_Z^2 \approx - 2 m_{H_u}^2 - \frac{\lambda^2}{\kappa^2}  \frac{A_\kappa^2}{4} \left( 1 - 4 z+  \sqrt{1- 8 z} \right) \, , 
\label{MZ2}
\end{align} 
with $z$ defined in Eq.~\eqref{eq:def_wz}. For the fine-tuning $\Delta$ we use the Barbieri-Giudice measure~\cite{Barbieri:1987fn} 
\begin{align}
\Delta\equiv \max_i\left\lbrace\Delta_{\lambda_i}\right\rbrace \,, \quad \Delta_{\lambda_i} \equiv \frac{\partial \log M_Z^2}{\partial \log
\lambda_i^2} \, ,
\label{deftuning}
\end{align}
where the maximum is taken over all UV parameters $\lambda_i$. For a thorough discussion of the tuning measure in extended GMSB models see e.g. Ref.~\cite{Evans:2013kxa}.
\section{The DGS Model}
\label{sec:DGS}
The field content of the DGS model~\cite{Delgado:2007rz} consists of the NMSSM fields (the MSSM fields plus a gauge singlet $S$), 
in addition to two copies of messengers in ${\bf 5} + {\bf \overline{5}}$ representations of SU(5). The superpotential is given by the NMSSM (see
Appendix for our conventions), the spurion-messenger couplings of ordinary gauge mediation and new singlet-messenger couplings. Apart from the NMSSM
part we have 
\begin{align}
W_{\rm DGS} & = X  \sum_{i=1}^2 \left(  \Phi_{u}^{(i)} \Phi_{d}^{(i)} +  \Phi_{T}^{(i)} \Phi_{\overline{T}}^{(i)} \right) \nonumber \\
& + S \left( \xi_D \Phi_{u}^{(1)} \Phi_{d}^{(2)}  +  \xi_T \Phi_{T}^{(1)} \Phi_{\overline{T}}^{(2)}   \right) \, , 
\end{align}
where $\Phi_u + \Phi_d$ and $\Phi_T+ \Phi_{\overline{T}}$ denote the doublet and triplet components in ${\bf 5}+{\bf \overline{5}}$, respectively, and
$X$ denotes the SUSY breaking spurion that takes the VEV $\langle X \rangle = M + F \theta^2$. This superpotential gives rise to soft SUSY breaking terms that can be
found in the Appendix. These are determined by six parameters:  the messenger scale $M$, the NMSSM couplings $\lambda$ and $\kappa$, the DGS couplings
$\xi_D$ and $\xi_T$ and the effective scale of soft SUSY breaking terms $\tilde{m} \equiv 1/(16 \pi^2) F/M$. One of these parameters (following DGS we choose
$\kappa$) can be eliminated by requiring correct EWSB. We also impose a unification condition for $\xi_D$
and $\xi_T$ that allows to eliminate one additional parameter, 
\begin{align}
\xi_D(M_{\rm GUT}) = \xi_T(M_{\rm GUT}) = \xi \, .
\end{align}
In the following analysis we will always assume this relation, but we have checked that the more general case of independent $\xi_D$ and
$\xi_T$ leads to similar phenomenology.

For heavy singlet-like scalars, all sparticles must be very heavy in order to satisfy the Higgs mass constraint in this model, and fall outside the discovery reach of the LHC~\cite{Delgado:2007rz}. It
was found in Ref.~\cite{Allanach:2015cia} that a significantly lighter SUSY spectrum is possible in the presence of a light singlet that pushes up the
Higgs mass via Higgs-singlet mixing. Having the light singlet mass fixed around 90$\div$100 GeV (where the constraints on the Higgs-singlet mixing from LEP~\cite{Schael:2006cr} are weakest) implies $\xi\sim10^{-2}$, while the Higgs-singlet mixing maximizing the push-up effect on the Higgs mass requires
$\lambda\sim10^{-2}$. For such small values of the couplings one has $m_S^2\approx - 16\tilde{m}^2 \xi_T^2 g_3^2$, so that the condition in Eq.~\eqref{eq:Svev_cond} for
correct EWSB is always fulfilled. As a result, the DGS model with a light singlet has $w\gg1$, so that $\tan\beta$ is approximately given by
Eq.~\eqref{eq:tanbeta_largew}. An interesting prediction of the DGS model with a light singlet is a singlino NLSP  with mass about $m_{\tilde{N}_1} \approx100$~GeV that decays to gravitino LSP and $a_1$ or $h_1$, which typically  decay further to $b\bar{b}$. An estimate for the decay length up to ${\cal O}(1)$ factors is given by 
\begin{equation}
c \tau_{\tilde{N}_1} \approx 2.5 \, {\rm cm} \, \left( \frac{100 \, {\rm GeV}}{M_{\tilde{N}_1}} \right)^5 \left( \frac{M} {10^6 \, {\rm GeV}}\right)^2 \left( \frac{\tilde{m}}{{\rm TeV}} \right)^2 \, . 
\end{equation}
Therefore the singlino is long-lived and results in displaced vertices within the LHC detectors for sufficiently low values of the messenger scale $M\lesssim10^7$~GeV, corresponding to an NLSP decay length $c \tau_{\tilde N_1}\lesssim{\mathcal O}(10)\ \rm{m}$. However, in order to see such a long-lived singlino  at the LHC, it must have been produced from the decay of  a heavy SUSY particle like gluino or squark, since the direct  production of a singlino is negligible due its small couplings. Therefore displaced signatures require a sufficiently light colored spectrum, which might be in conflict with existing LHC searches. 

\begin{figure}
\centering
\includegraphics[width=0.45\textwidth]{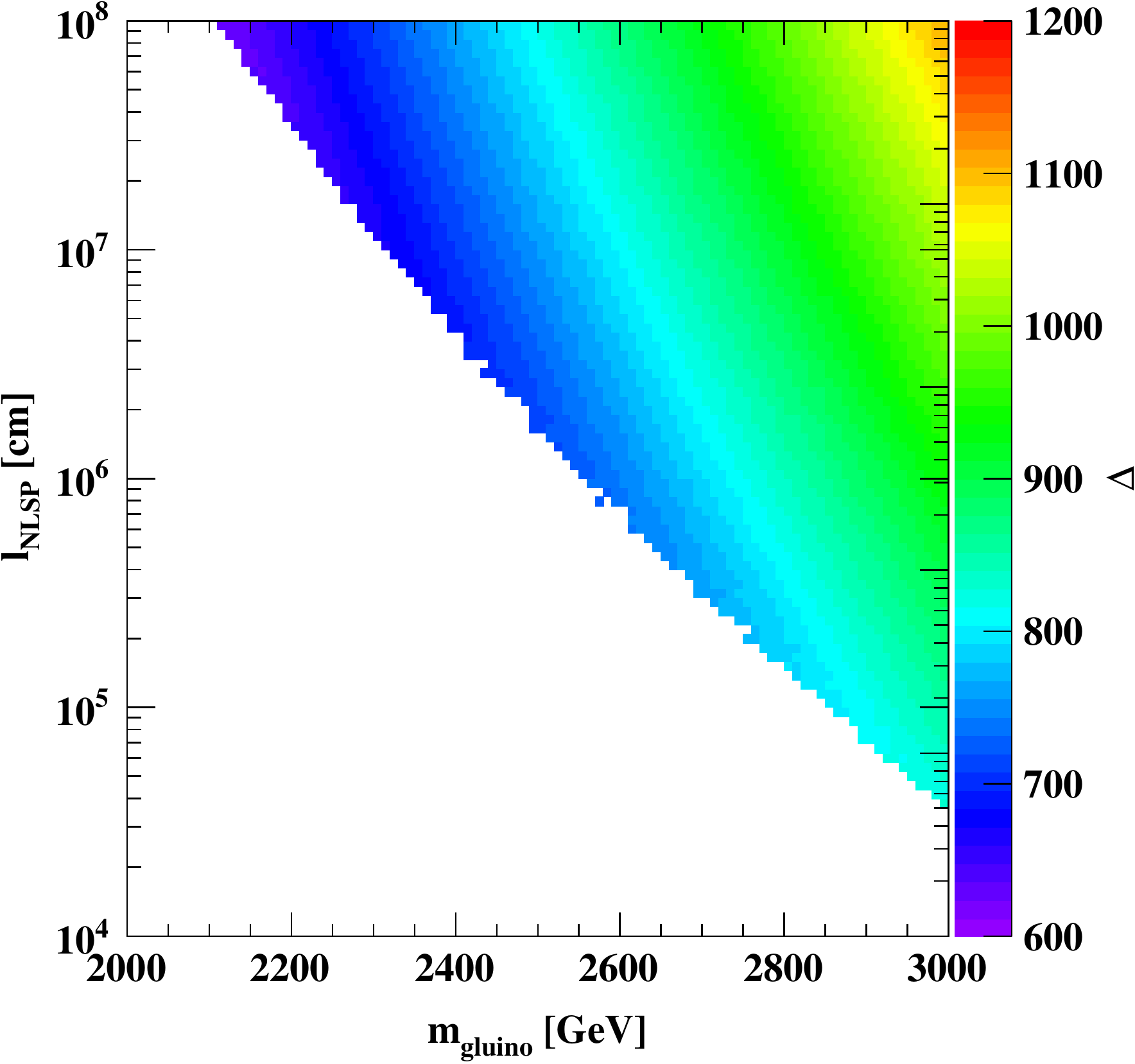}
\caption{Gluino mass vs. NLSP decay length $c \tau_{\tilde N_1}$ in the DGS model after applying all  current LHC constraints. The colour map indicates the values of the fine-tuning measure $\Delta$. \label{fig:DGS_mgluino_lNLSP}}
\end{figure}

In order to assess this issue, we have updated the analysis of the DGS model in Ref.~\cite{Allanach:2015cia} using the latest LHC constraints. For this analysis we have used the public code {\tt NMSSMTools}~\cite{Ellwanger:2004xm,Ellwanger:2005dv}, which computes the SUSY and Higgs spectrum and checks the latest LHC constraints in the Higgs sector (LHC constraints on SUSY sector will be dealt with later with {\tt CheckMATE2}). We have performed large MCMC scans of the parameter space of the DGS model and sorted the results in a 2D histogram in the plane of the gluino mass vs. the decay length of the singlino NLSP. In each bin of this histogram, we have kept the point in parameter space minimising the fine-tuning as defined in Eq. (11). In Fig.~\ref{fig:DGS_mgluino_lNLSP}  we present a map of fine-tuning in the plane of the gluino mass and the decay length of the singlino NLSP. We see that for NLSP decay lengths corresponding to a displaced vertex, the gluino mass is pushed far beyond 3~TeV.  Thus, the DGS 
model does not predict any displaced signatures that could be observed at the LHC. We also note that even for larger NLSP decay length, for which the NLSP is stable from the LHC perspective, the gluino must be heavy and may be beyond the LHC discovery reach. Numerically, we find a lower bound on the gluino mass of $1.8 \, {\rm TeV}$, and a  lower bound on the tuning of $\Delta \ge 600 $.

These negative conclusions can be traced back to recent heavy Higgs searches in the $\tau\tau$ channel \cite{Aaboud:2017sjh}, which exclude previously viable points with light gluinos. This happens because successful EWSB  in the DGS model with a light singlet requires rather large values of $\tan\beta\gtrsim30$,  which pushes the heavy MSSM-like Higgs bosons to values above 1.3 TeV \cite{Aaboud:2017sjh}, and thus requires larger SUSY scales.

Therefore, in order to have a light sparticle spectrum that can be tested at the LHC, a  model of NMSSM + GMSB is desirable that
can provide sufficiently heavy MSSM-like Higgs bosons.
In the DGS model the gluino mass is correlated with $m_{a2}$ since the scale of both parameters is set by $\tilde{m}$. In order to avoid stringent lower
bounds on the gluino mass, one should look for a model in which  the correlation between gluino mass and $m_{a2}$ is broken by new contributions to the UV soft masses and/or $w\approx1/3$ (for which $\tan\beta$ is generically smaller so the LHC constraints on $m_{a2}$ are weaker).  In the following sections we present two models that satisfy these requirements and therefore simultaneously allow both for a gluino much lighter than in the DGS model and sufficiently small singlino decay lengths to have displaced vertices at the LHC.

\section{The U Model}
\label{Umodel}
In this new model, we take one copy of messengers and add all possible linear couplings of the messenger $\Phi_{u}$:
\begin{align}
W_{\rm U} & = X \left(  \Phi_{u} \Phi_{d}  +  \Phi_{T} \Phi_{\overline{T}}  \right)  \nonumber \\
& + \lu  Q_3 U_3 \Phi_{u} +  \lsd S  \Phi_{u} H_d \, .
\end{align}
This gives rise to soft SUSY breaking terms that can be found in the Appendix, and are now determined by six parameters:  $\tilde{m},
M,\lambda, \kappa$, $\lambda_t$ and $\lambda_{S_d}$. 
Note that one might impose the condition $\lambda_{S_d} y_t = \lambda_t \lambda$ that would result from explicit messenger-Higgs
mixing~\cite{Evans:2011bea}. However, we found that for small values of $\lambda\lesssim10^{-2}$ (as required to avoid experimental constraints on the Higgs-singlet
mixing), $\lambda_{S_d}$ would be too small to allow for $m_{h_1}$ to be in the preferred window between about 90 and 100 GeV in order to to significantly enhance the Higgs
mass. Thus, in the following we assume that $\lambda_{S_d}$ and $\lambda_t$ are independent parameters, so that there is one additional parameter as compared to the DGS model.

\begin{figure}
\centering
\includegraphics[width=0.45\textwidth]{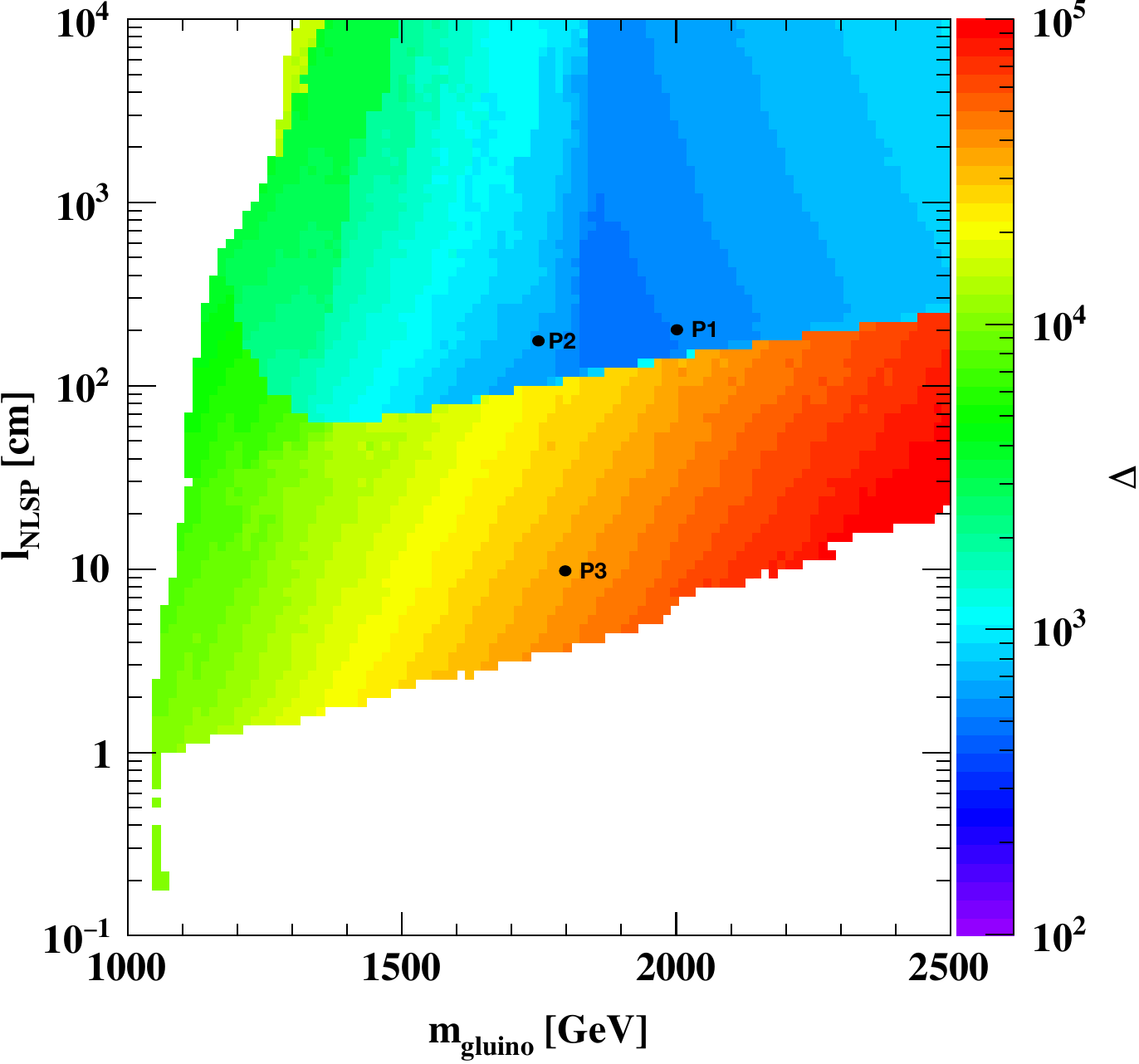}
\caption{The same as in Fig.~\ref{fig:DGS_mgluino_lNLSP} but for the U model and without applying constraints from LHC sparticle searches.  Also shown are the benchmark points P1-P3 of Table~\ref{tab:benchmarks}.  
\label{fig:Umodel_mgluino_lNLSP}}
\end{figure}

In Fig.~\ref{fig:Umodel_mgluino_lNLSP} we present a map of fine-tuning  in the plane of the gluino mass and the
decay length of the singlino NLSP, obtained analogously to Fig.~\ref{fig:DGS_mgluino_lNLSP}. One can clearly distinguish two different regions here, in the upper part  the ``DGS-like" region, that is characterised by a large NLSP decay length $c \tau_{\tilde{N}_1} \gtrsim 100 \,  {\rm cm}$, small fine-tuning $\Delta \lesssim 10^3$, a light singlet spectrum ($m_{a_1} \approx 30 \div 40 \GeV$, $m_{h1} \approx 90 \GeV$, $m_{\tilde{S}} \approx 100 \GeV$) and an essentially constant input value $\lambda_{S_d} \sim 10^{-2}$. Representative for the DGS-like region are the benchmark points P1 and P2 in Table~\ref{tab:benchmarks}. A closer look reveals that this region actually falls into two sub-regions, which can be separated  by the value of the gluino mass as visible in Fig.~\ref{fig:Umodel_mgluino_lNLSP}. For gluino masses above 1.8 TeV, the value of $|\lu|$ is essentially constant, $\sim 0.3$, while it increases up to $\sim0.7$ towards lighter masses. Moreover, these parts are distinguished by the value of $\
tan \beta$, which  is about 10  in the left part and about 20 in the heavy gluino part. The latter is represented by P1 while P2 
exemplifies the light gluino part of the DGS region. 

The other region with smaller NLSP decay lengths $c \tau_{\tilde{N}_1} \lesssim 100 \,  {\rm cm}$ is represented  by benchmark P3 in Table~\ref{tab:benchmarks}. 
This region features much larger tuning $\Delta \gg 10^3$, essentially constant $\tan \beta \approx 10$, a heavier singlet spectrum ($m_{a1}, m_{\tilde{N}_1} \gtrsim 200 \GeV$), large constant values of $|\lambda_t| \approx 0.7$ and $\lambda_{S_d} \approx 0.2$ and a potentially lighter gluino mass compared to the DGS-like region. From Fig.~\ref{fig:Umodel_mgluino_lNLSP} one can see that a singlino decay length $\mathcal{O}(1) \, {\rm cm}$ is possible for a gluino as light as about 1~TeV, without being  in conflict with Higgs sector constraints. Nevertheless, such a light gluino might be already excluded by direct searches at the LHC. 

We have therefore taken into account the LHC limits from direct SUSY searches with {\tt CheckMATE2}~\cite{Dercks:2016npn}. 
To briefly summarise the workflow, {\tt CheckMATE2} uses {\tt Pythia8}~\cite{Sjostrand:2014zea} to generate all accessible $2\rightarrow 2$ processes
followed by a detector simulation with {\tt Delphes3}~\cite{deFavereau:2013fsa}.  Variables and cuts used in experimental analyses are then
implemented as closely as possible to ``recast'' the analysis, and the expected number of signal events that pass the cuts are validated against
published benchmarks and cut flows.  This validated analysis can then be used to test New Physics models against published upper limits. In the
presence of multiple signal regions that may potentially be sensitive to model predictions, {\tt CheckMATE2} selects only the most sensitive region
with respect to the expected background.

\begin{figure}
\centering
\includegraphics[width=0.45\textwidth]{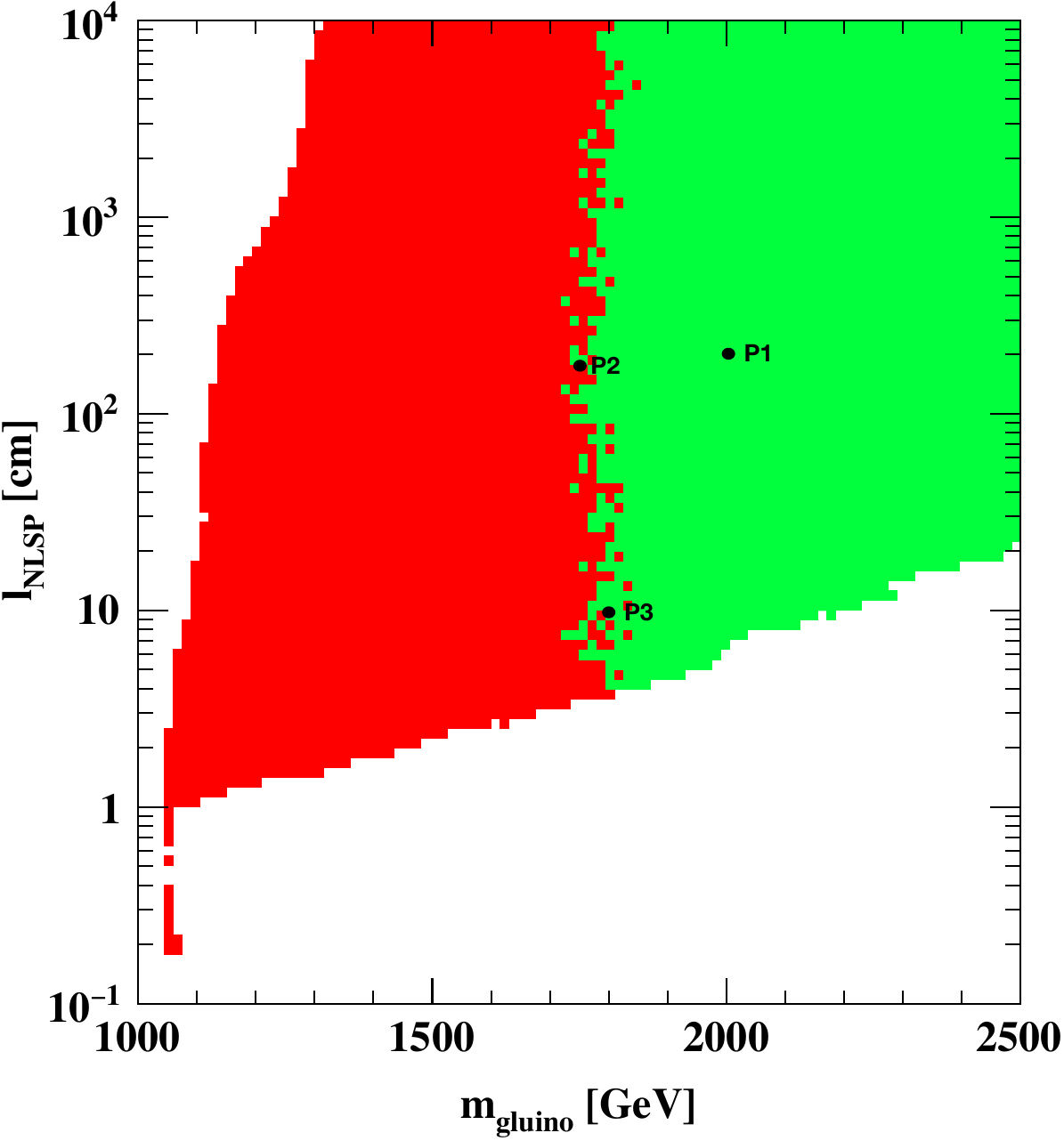}
\caption{Exclusion limits  in the U model obtained with {\tt CheckMATE2}. Red (green) points are excluded (allowed) by direct SUSY searches at the LHC. Also shown are the benchmark points P1-P3 of Table 1.  \label{fig:Umodel_excl}}
\end{figure}

We summarize the results in Fig.~\ref{fig:Umodel_excl} in the same parameter space as in Fig.~\ref{fig:Umodel_mgluino_lNLSP},
where red (green) points are excluded (allowed) by current LHC constraints. This plot shows that direct LHC searches still allow for a gluino as light
as about 1.7~TeV, for essentially any NLSP decay length. The bound is significantly weaker than in typical simplified models presented by the experimental
collaborations. The main reason for the relaxed constraints is that the wino is lighter than the left-handed sleptons of the first two generations, and therefore dominantly decays to
the lightest stau (which has some non-negligible left-handed component), resulting in $\tau$'s instead of leptons in the final state. This feature strongly relaxes the mass
limits both for direct production of winos, as well as production of gluinos decaying predominantly to winos. We found that the most constraining
searches for this model are the ATLAS searches with two same-sign leptons or tri-leptons~\cite{Aaboud:2017dmy}, and jets and missing energy~\cite{Aaboud:2017vwy}.

\begin{table*}
\centering
\begin{tabular*}{\textwidth}{@{\extracolsep{\fill}}|c|ccc|ccc|}
\hline
   & P1 & P2 & P3 & P4 & P5 & P6 \\
\hline
$\tilde m$~[TeV] & $1.7$ & $1.5$  & $1.5$ & $0.87$ & $1.0$& $1.0$ \\
 $M$ & $2.8 \times 10^{6}$& $3.1 \times 10^{6}$ & $2.5 \times 10^{6}$ & $5.6 \times 10^{6}$ & $5.1 \times 10^{6}$ & $1.6 \times 10^{6}$\\
  $\lambda$ & $4.6 \times 10^{-3}$ & $4.4 \times 10^{-3}$  & $ 1.1 \times 10^{-3}$ & $4.9 \times 10^{-3}$ & $5.4 \times 10^{-3}$ & $2.5 \times 10^{-3}$  \\
  $\kappa$ & $1.4 \times 10^{-4}$ & $1.2 \times 10^{-4}$ & $ 4.3 \times 10^{-5}$ & $1.5 \times 10^{-4}$ & $2.1 \times 10^{-4}$ & $6.5 \times 10^{-5}$ \\
 $\lambda_t$ & $-0.33$ & $-0.48 $ & $-0.73$ & $-0.64 $ & $- 0.38 $& $0.76 $\\
 $\lambda_{S_d}$ & $0.022 $ & $0.028 $ & $ 0.17 $& - & - & - \\
 $\xi$ & - & - & - & $ 0.012 $ & $ 0.010$ & $  0.010$\\
  $\tan \beta$   & 18 & 11 & 10 & 9.1 & 17 & 8.7  \\
  \hline
$m_{\tilde g}$~[TeV] & 2.0 & 1.7 & 1.8 & 2.0 & 2.3 & 2.2 \\
$m_{\tilde d_R}$~[TeV] & 2.6 & 2.2 & 2.3 & 2.0 & 2.4 & 2.3\\
$m_{\tilde t_1}$~[TeV]  & 2.1 & 1.7 & 2.2 & 1.8 & 2.0 & 2.2  \\
$m_{\tilde N_1}$~[GeV] & 95 & 96 & 200 & 96 & 97 & 106  \\
$m_{\tilde N_2}$~[GeV]  & 370 & 320 & 330 & 380 & 440& 430 \\
$m_{a_1}$~[GeV]  & 26 & 32 & 290 & 26 & 24 & 26 \\
$m_{a_2}$~[TeV] & 1.7 & 1.9 & 2.7 & 1.6 & 1.4 & 2.1\\
$m_{h_1}$ [GeV]   & 89 & 89 & 110 & 89 & 91& 101 \\
$m_{\mu L}$ [GeV]   & 960 & 830 & 820 & 710 & 840& 790 \\
$m_{\mu R}$ [GeV]   & 480 & 430 & 520  & 390 & 420 & 450\\
$m_{\chi_1^\pm}$ [GeV]   & 720 & 620 & 640  & 720 & 840 & 830\\
\hline
$c \tau_{\tilde N_1}$ [cm] & $200$ & $200 $ & $10 $ & $200  $ & $200 $  & $12$ \\
$\Delta$ & $530 $ & $680 $ & $37000 $ & $440 $ & $310$& $720$ \\
\hline
\end{tabular*}
\caption{Benchmarks consistent with all experimental constraints, including LHC direct search limits. P1-P3 are points in the  U model, while P4-P6 belong to the DGSU model. All points have reduced SM-like Higgs signal strengths of about 0.84, corresponding to a Higgs-singlet mixing angle of $\cos \theta \approx 0.92$, while the signal strengths of the singlet-like state are about 0.16.
\label{tab:benchmarks}}
\end{table*}

\begin{figure}
\centering
\includegraphics[width=0.5\textwidth]{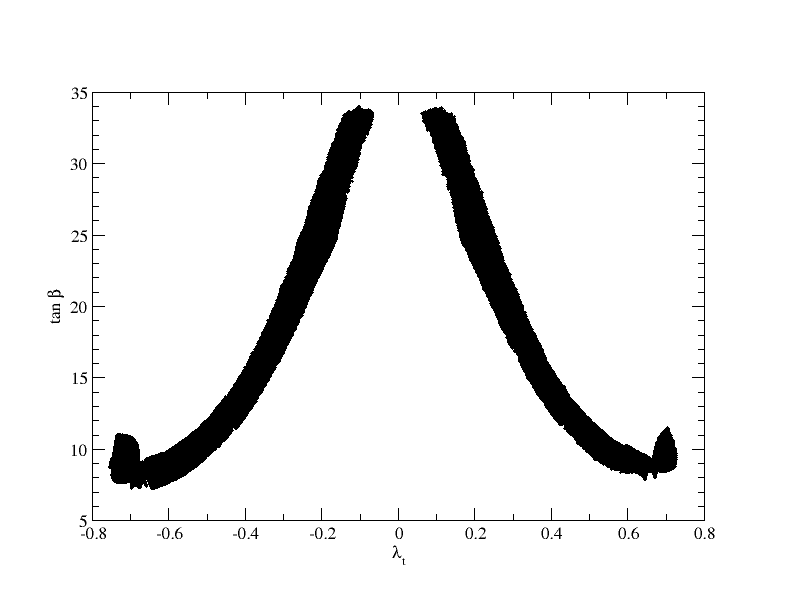}
\caption{Correlation of $\tan\beta$ and $\lu$ for points with $m_{\tilde{g}} < 2.5$~TeV and $c \tau_{\mathrm{\tilde{N}_1}} < 10^4$~cm displayed in Figs.~2 and 3.\label{fig:Umodel_tanb_lt}}
\end{figure}

Let us now discuss how such a light gluino  in the U model can be compatible with the constraints from the SM-like Higgs mass and experimental searches for heavy Higgses.  
First of all, the points with a light gluino feature large $\lambda_t$, which implies large $A_t$, so that the Higgs mass is enhanced not
only by the push-up effect but also by the loop contribution from stop mixing. However, we recall that in the DGS model the lower bound on the SUSY scale
does not arise from the Higgs mass constraint, but from direct searches for heavy MSSM-like Higgs bosons. In the U model instead, the pseudoscalar Higgs mass $m_{a_2}$ is enhanced by large
$\lambda_t$, as a result of the contribution to the soft Higgs mass parameter $m_{H_u}^2\sim-9y_t^2\lambda_t^2\tilde{m}^2$, cf.~\eqref{m2U}. Moreover, larger values of
$\lambda_t$ are correlated with smaller $\tan\beta$, as shown in Fig.~\ref{fig:Umodel_tanb_lt}. This is because larger $\lu$ results in smaller $w$, which
follows from
\begin{equation}
 z=\frac{m_S^2}{A_\kappa^2}\approx \frac{1}{6\lsd^2}\left(\lu^2-g_2^2\right) \, , 
\end{equation}
and Eq.~\eqref{eq:def_wz}. For $\lu\gtrsim0.7$, $w$ approaches $1/3$ for which $\tan\beta$ is twice as small as in the limit $w\gg1$, as
explained\footnote{For large $\lu$, $\tan\beta$ is suppressed even more due to the RG effect of $A_t$ which makes 
$A_\lambda$ positive at the EW scale and cannot be neglected in Eq.~\eqref{eq:tanbeta_w13}. This effect is partially compensated by the fact
that $\lambda_t$ increases $m_{a2}$ that enters in the numerator of the formula for $\tan\beta$ in Eq.~\eqref{eq:tanbeta_w13}. }
 in Section~\ref{sec:EWSB}. Therefore, the combined effect of increased $m_{a2}$ and decreased $\tan\beta$ renders the LHC searches for heavy Higgs bosons with  $\tau\tau$ final states essentially insensitive to the U model with large $\lambda_t$ and light gluinos. 

We however notice that the points with light gluinos are more fine-tuned than DGS-like solutions with smaller $\lu$ and gluinos above 1.8 TeV. This is because the tuning associated with $\lambda_t$ dominates the total tuning in almost all points of the U model. Using the approximate expression for $M_Z^2$ in Eq.~\eqref{MZ2}, one can easily show that the $\lambda_t$ contribution to the tuning $\Delta_{\lu}$ is approximately given by 
\begin{align}
\Delta_{\lambda_t} & \approx 6 \lambda_t^2\frac{\tilde{m}^2}{M_Z^2} \left( 3 y_t^2 +  \frac{\lambda^2}{\kappa^2}  \lambda_{S_d}^2  \right)   \, ,
\label{tuningU}
\end{align}
where the first term is  the contribution to $m^2_{H_u}$, and the second is the contribution to the effective $\mu$-term. 

 While $\lambda^2/\kappa^2 \sim 10^3$ is essentially constant  in both regions, the smallness of $\lambda_{S_d} \sim 10^{-2}$ in the DGS-like region implies that the second term in Eq.~\eqref{tuningU} is small, and the tuning is essentially determined by $\lambda_t^2 \tilde{m}^2$, with a minimal value of $\Delta \approx 450$. Instead the other region features large $\lambda_{S_d} \sim 0.2$, which together with large $\lambda_t$ leads to a significant enhancement of the tuning with respect to the DGS-like region, by about a factor 100.  This estimate explains the large increase in the tuning when crossing between the two different regions in Fig.~\ref{fig:Umodel_mgluino_lNLSP}. It also accounts for the approximately constant tuning over the two DGS-like sub-regions, where a decrease in the gluino mass (i.e. $\tilde{m}$) is compensated by an increase in $\lambda_t$, such that the total tuning remains essentially constant, up to ${\cal O}(1)$ factors. 

We conclude this section with a discussion of the three  benchmark points P1-P3 in Table~\ref{tab:benchmarks}, which represent sample spectra of the U
model with a light SUSY spectrum compatible with all LHC constraints, with a singlino decay length roughly below two meters. The point P1 features a
tuning of $\Delta \approx 500$, with a  gluino mass around 2 TeV and NLSP singlino decay length about 2~m. This point represents the DGS-like region
in the upper right corner with large $\tan \beta$.
The point P2 has larger $\lambda_t\approx0.5$, which allows for a lighter gluino, although the tuning slightly increases as discussed above. The larger $\lambda_t$ implies smaller $\tan \beta \approx 11$, while the light singlet spectrum is quite similar to P1, with a NLSP singlino decay length that is still relatively large, $c \tau_{\tilde{N}_1} \approx  1.9 \, {\rm m}$. For both P1 and P2, the singlino NLSP mainly decays to the gravitino and the lightest pseudoscalar $a_1$ (with mass $\sim 30$~GeV) which in turns decays promptly to $b \overline{b}$. Characteristic for these benchmarks is a light Higgs state $h_1$ with mass around 90 GeV, which is mainly a SM singlet with a small doublet admixture corresponding to a mixing angle $\cos^2 \theta \approx 0.84$. This state might account for the $2\sigma$  excess observed at LEP~\cite{Schael:2006cr,Belanger:2012tt} in the $b\bar{b}$ channel, and could also explain the tentative hint for a light Higgs decaying into diphotons at CMS~\cite{CMS:2017yta}, although 
the signal strength of the light singlet state in the diphoton channel is about a factor of three smaller than the current central value of the CMS excess. 

 On the other hand, NLSP decay lengths below $1 \, {\rm m}$ can only be obtained for the price of increased fine-tuning, in the region where $\lambda_t \approx 0.7$ is constant. This is exemplified by the point P3, which features $c \tau_{\tilde N_1}\approx10$~cm and a tuning of at least $\Delta \approx 10^4$. These points have a heavier Higgs spectrum, and in particular the lightest pseudoscalar is now heavier than 250 GeV, while the NLSP singlino is above 200 GeV. In contrast to P1 and P2, here the  singlino mainly decays to gravitino and $h_1$ (with mass $\sim 110$~GeV), since $a_1$ is too heavy. 
 
\section{The DGSU Model}
\label{DGSUmodel}
Motivated by the previous analysis, we finally consider a model that combines the virtues of the DGS and the U model, thus allowing for light gluinos, small singlino decay lengths and low tuning.  In this scenario, dubbed the ``DGSU model", we take two messenger copies (i.e. $N=2$) and introduce the following couplings:
\begin{align}
W_{\rm DGSU} & = S \left( \xi_D \Phi_{u}^{(1)} \Phi_{d}^{(2)}  +  \xi_T \Phi_{T}^{(1)} \Phi_{\overline{T}}^{(2)}   \right) \nonumber \\
&   + \lu  Q_3 U_3 \Phi_{u}^{(2)} +  \lsd S  \Phi_{u}^{(2)} H_d \, .
\label{WU}
\end{align}
We impose the DGS unification condition at the GUT scale
\begin{align}
\xi_D(M_{\rm GUT}) = \xi_T(M_{\rm GUT}) = \xi \, .
\end{align}
In contrast to the U model we can\footnote{This is because the singlet scalar mass is now set by $\xi\sim10^{-2}$ as in the DGS model, instead of $\lambda_{S_d}$ as in the U model.}  now impose  the 
Higgs-messenger mixing condition (for simplicity at the messenger scale)
\begin{align}
\lambda_{S_d} (M) y_t (M) = \lambda_t (M) \lambda (M) \, .
\end{align}
The superpotential couplings lead to additional contributions that
can be found in the Appendix, and are given in terms of the six independent parameters $\tilde{m}$, $M$, $\lambda$, $\kappa$, $\xi$ and $\lambda_t$
(the same number as in the U model).

\begin{figure}
\centering
\includegraphics[width=0.5\textwidth]{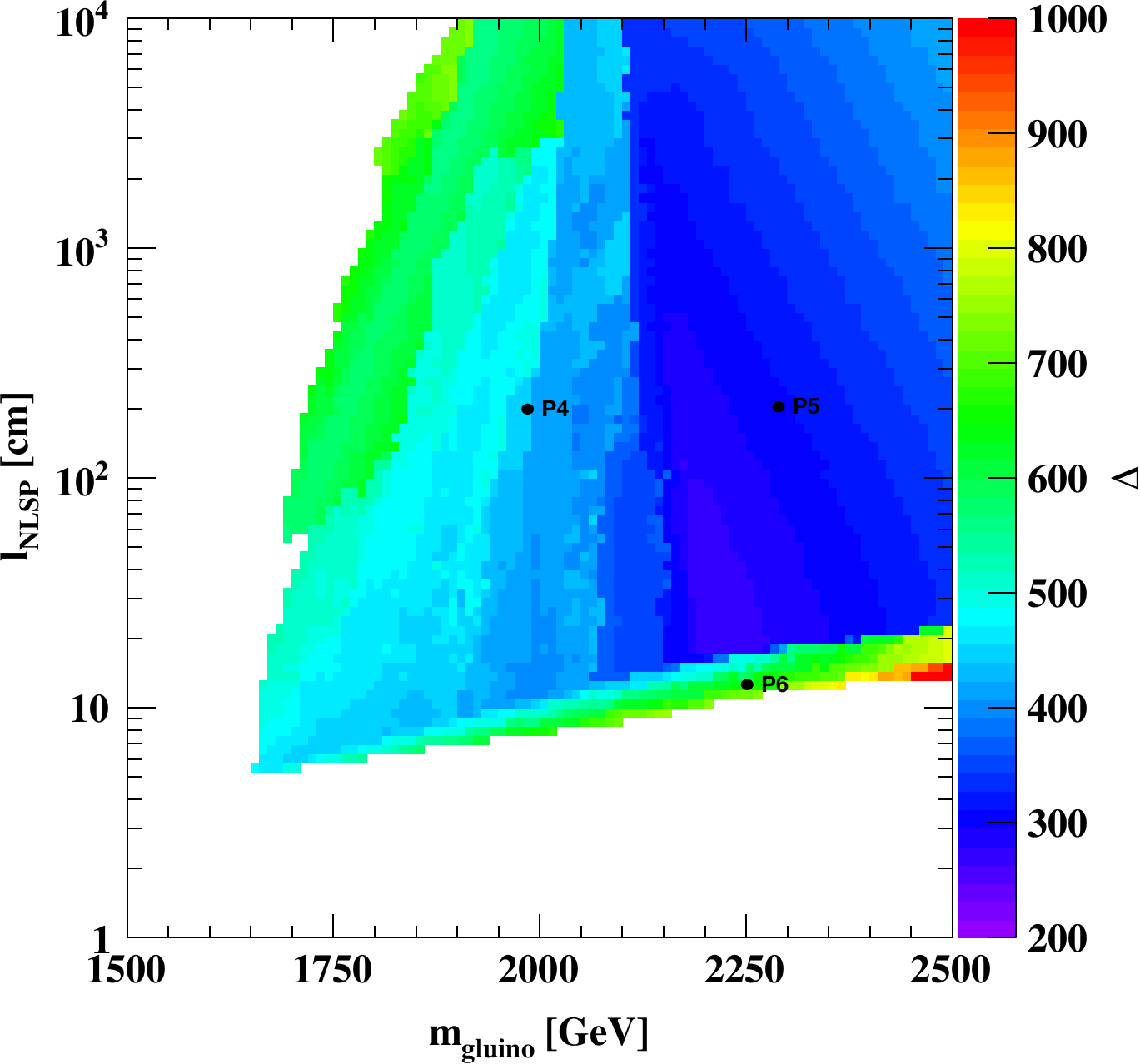}
\caption{The same as in Figs. 1 and 2, but for the DGSU model and without applying constraints from LHC sparticle searches. Also shown are the benchmark  P4--P6  of Table~\ref{tab:benchmarks}.  } \label{fig:DGSU_mgluino_lNLSP}
\end{figure}

\begin{figure}
\centering
\includegraphics[width=0.45\textwidth]{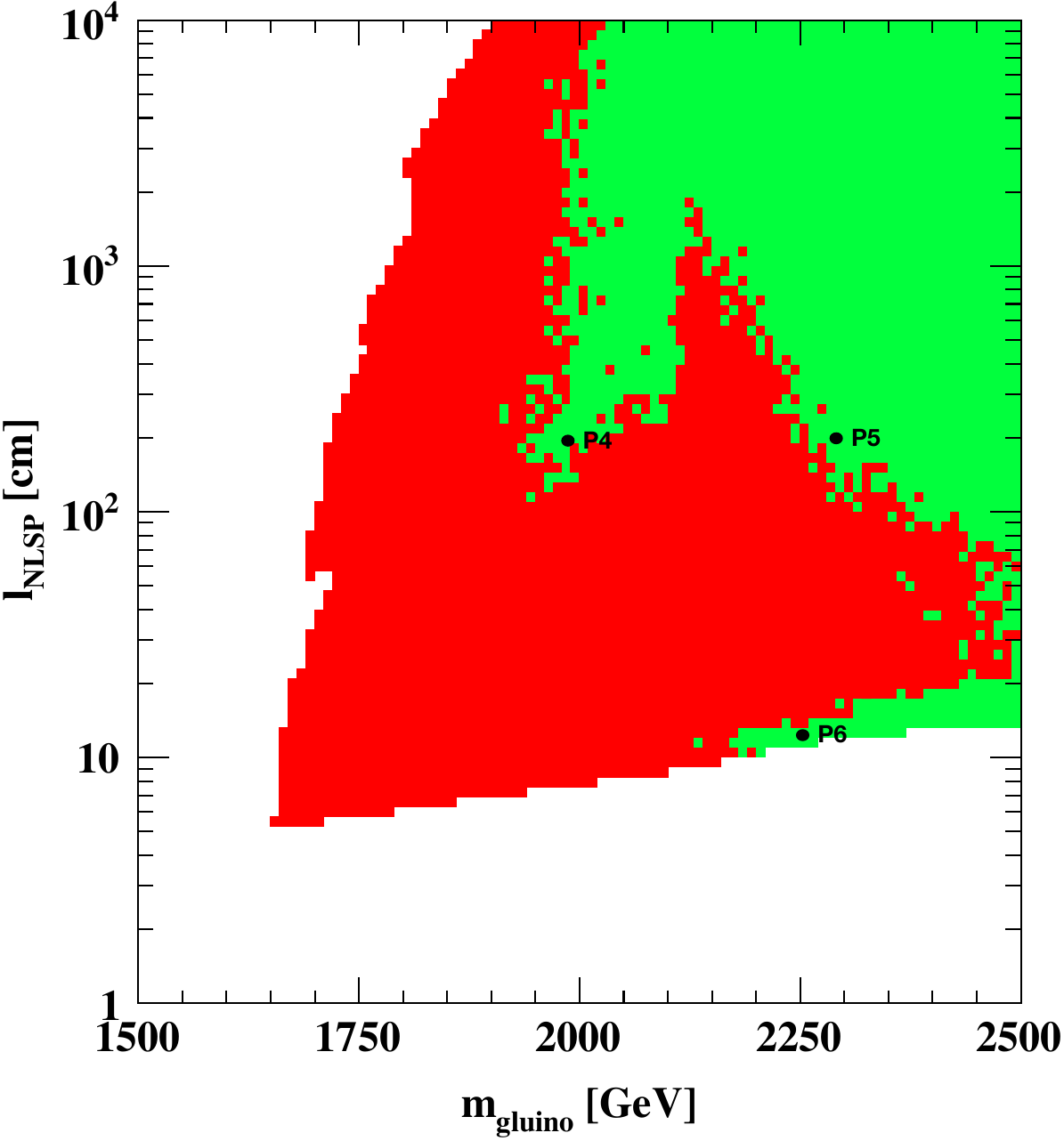}
\caption{Exclusion limits  in the DGSU model obtained with {\tt CheckMATE2}. Red (green) points are excluded (allowed) by direct SUSY searches at the LHC. Also shown are the benchmark points P4-P6 of Table 1.  \label{fig:DGSUmodel_excl}}
\end{figure}

In Fig.~\ref{fig:DGSU_mgluino_lNLSP} we present a map of fine-tuning in the plane of the gluino mass and the decay length of the singlino NLSP, obtained analogously to Figs. 1 and 2. However, comparing to the exclusion limits from direct LHC searches shown in Fig.~\ref{fig:DGSUmodel_excl}, we see that they are stronger
than in the U model and the allowed points can have gluino masses only slightly below 2 TeV. The reason for this is that in the DGSU model there are
two messengers (in contrast to one in the U model). Since the minimal gauge mediation contribution to gaugino masses is proportional to the number of
messengers $N$ while the corresponding contribution to sfermion masses scales as $\sqrt{N}$, cf.~\eqref{softMGM}, this makes sfermions lighter for a
given gaugino masses. As a result, the squarks of the first two generations are comparable to the gluino mass and their production cross-section is
non-negligible. For the same reason left-handed sleptons of the first two generations can now be lighter than wino, which results in more leptons in
the final state (instead of $\tau$'s as in the U model). The latter feature of the DGSU model also explains why the constraints for larger NLSP decay
lengths are weaker, since the larger decay length is obtained for larger messenger scale, which in turn results in heavier sleptons (as compared to
gauginos) due to their stronger renormalization by electroweak gauginos. Nevertheless, a gluino mass of 2 TeV is still viable for a decay length
$\mathcal{O}(1)$~m, while decay lengths roughly between  20 and 100~cm implies a limit for the gluino mass of almost 2.5~TeV. We should also emphasize that
this limit is particularly strong not only due to gluino/squark production, but also due to direct production of sleptons and winos, whose  mass is correlated
with the gluino mass. Therefore also direct searches for direct electroweak production set important constraints~\cite{Sirunyan:2017lae}, besides analyses using jets and missing energy. 

Note that for decay lengths of  $\mathcal{O}(10)$~cm, there is  a small strip of allowed points (represented by benchmark point P6), which however correspond to larger tuning, cf. Fig.~\ref{fig:DGSU_mgluino_lNLSP}. This is a consequence of large values of $|\lambda_t|$, which also explains the relaxed LHC constraints. For large values of $|\lambda_t|$, there is a large positive contribution to stop masses at the messenger scale, cf.~\eqref{m2U}, which leads to heavier stop masses at low scales, so that gluino decays to stops are kinematically forbidden.

We notice that the DGSU model shares many features with the DGS model, or rather  the DGS-like region of the U model. In particular, all points (represented by benchmarks P4-P6 in Table~\ref{tab:benchmarks}) have a well-defined
singlet sector with a singlet-like scalar $h_1$ around 90--100~GeV, a singlet-like pseudoscalar $a_1$
around 20--30~GeV and a singlino NLSP between 90--100 GeV. The singlino NLSP decay length can be rather short, of the order of ${\cal O}(10)$~cm, and it decays mainly to gravitino and $a_1$, which decays promptly to $b \overline{b}$.
The input parameters are essentially constant throughout the whole region, with $\lambda \sim 10^{-3}$ and $\xi \sim 10^{-2}$. As in the DGS-like region of the U model one can further distinguish two sub-regions, which are characterised by different values of $|\lambda_t|$ (that is essentially constant $\sim 0.4$ for gluino masses above $\sim$ 2.2~TeV, and below this value starts increasing towards lighter gluino masses, up to $\sim 0.8$) and $\tan \beta$, which is correlated with $\lambda_t$ similar as in Figure~\ref{fig:Umodel_tanb_lt}. These two parameters essentially control the mass of the MSSM-like pseudoscalar $a_2$ and the tuning. The mass of the former  grows for larger $\lambda_t$ (a feature inherited from the U model), and thus is no longer correlated with the gluino mass so that $H/A\to\tau\tau$ searches do not  constrain this model. Moreover, large values of  $\lambda_t$ lead to a large contribution to the Higgs mass from stop mixing, which implies that the overall SUSY scale can be lowered, and 
gluinos and stops can be 
quite light. The DGSU model is thus a perfect realization of the pNMSSM scenario analysed in Ref.~\cite{Allanach:2016pam}, and motivates the combined searches using displaced and prompt signatures advocated in that article. 

Finally, the DGSU model is also much less fine-tuned than the points with light gluinos in the U model. This can be again understood from the fact that $\lambda_t$ (which is sizable) dominates the total tuning, and contributes as in Eq.~\eqref{tuningU} with $\lambda_{S_d}$ replaced by $\xi$. Similar to the DGS-like region of the U model, this contribution is small, so that the tuning is controlled by $\lambda_t^2 \tilde{m}^2$, with a minimal\footnote{This is a factor of a few smaller than the minimal tuning found for a broad class of extended GMSB models in the context of the MSSM~\cite{Evans:2013kxa}.}  value of $\Delta \approx 300$, which is reached for intermediate values for the gluino mass around 2.3~TeV, where the product of $\lambda_t$ and $m_{\tilde{g}}$ is minimal.

\section{Summary and Conclusions}
\label{conclusions}
In this article we have analyzed models of extended Gauge Mediation in the context of the NMSSM.  The simplest scenario (DGS) requires rather heavy gluinos, which are constrained not by direct SUSY searches but rather a two-fold restriction coming from the Higgs sector. On the one hand, the mass of the SM-like Higgs requires sizable loop corrections, on the other hand direct LHC Higgs searches in the $\tau \tau$ channel set stringent lower limits on the mass of the heavy Higgses, which in turn is tied to the SUSY scale.  We have proposed two new models with singlet/Higgs-messenger couplings (the U model and the DGSU model), which ease these constraints and thus allow for gluinos as light as allowed by direct searches. First, the presence of a direct coupling of stops to the messenger fields generate sizable $A_t$ at the messenger scale, which in turn allows for much lighter stops (and thus also other sparticles) consistent with the measured SM-like Higgs mass. Second, the same coupling also generates new 
contributions to the soft Higgs mass, which allows to both increase the MSSM-like Higgs doublet mass and reduce $\tan\beta$, such that the LHC
searches for heavy Higgs are satisfied without raising the overall SUSY scale. To identify the lower bound on the gluino mass in these models, we have
recasted the existing LHC searches and found that gluinos can be as light as about $1.7 \, {\rm TeV}$ (U model) and   $2.0 \, {\rm TeV}$ (DGSU model).
The tuning in both models is rather low, and can be as small as 2\permil (U model) and 3\permil (DGSU model). The phenomenology is quite different in
the two models: in the U model there are essentially two distinct regions, one with a DGS-like spectrum featuring a light singlet sector, large
singlino decay lengths roughly above a meter
 (but small enough to see a displaced vertex in the LHC detectors) and low fine-tuning. The other region is characterized by a heavy singlet sector, less displaced singlino decays with $c \tau_{\tilde N_1}\approx1\div10$~cm and  large tuning. The DGSU model combines the most interesting phenomenological features of both regions: light gluinos in the reach of LHC, a light singlet sector with a 98 GeV state (that might account for the LEP excess and improves the fit to the CMS data hinting at a new light state decaying to $\gamma\gamma$), and displaced singlino decays into $b \overline{b}$ + MET with decay lengths as small as a few cm. In Table~\ref{tab:benchmarks} we have collected benchmark points representative for the relevant parameter regions and the two models.

\section*{Acknowledgements}
The authors acknowledge the support of France-Grilles and the OCEVU Labex (ANR-11-LABX-0060) for providing computing resources on the French National Grid Infrastructure, and support from the French research project Defi InFIniti - AAP 2017.
MB and RZ thank the Galileo
Galilei Institute for Theoretical Physics and INFN for
hospitality and partial support during the completion
of this work. MB has been partially supported by the National Science Centre, Poland, under the research grant no. 2017/26/D/ST2/00225. 
\appendix
\section{Models}
In this Appendix we provide further details about the models discussed in the main text, including the full superpotential, its motivation by symmetries, and the complete list of soft terms (obtained using the results of Ref.~\cite{Evans:2013kxa}).   

\section*{0) NMSSM + Gauge Mediation}
\setcounter{equation}{0}

In order to fix notation, we define the NMSSM  and minimal gauge mediation by the superpotential
\begin{align}
\label{NMSSM}
W  & =  X   \sum_{i=1}^N \left(  \Phi_{u}^{(i)} \Phi_{d}^{(i)} +  \Phi_{T}^{(i)} \Phi_{\overline{T}}^{(i)} \right)  + \lambda S H_u H_d + \frac{\kappa}{3} S^3  \nonumber \\
& + Q^T \yu U H_u + Q^T \yd D H_d + L^T \ye E H_d  \, , 
\end{align}
where $X$ denotes the SUSY breaking spurion that takes the vev $\langle X \rangle = M + F \theta^2$, $N$ is the number of messengers in complete ${\bf 5}+{\bf \overline{5}}$ representations of SU(5), and $\Phi_u + \Phi_d$ and $\Phi_T+ \Phi_{\overline{T}}$  denote the doublet and triplet components in ${\bf 5}+{\bf \overline{5}}$, respectively. The spurion vev will induce soft masses and A-terms defined as
\begin{align}
- {\cal L} & =\tilde{q}_L^T \tilde{m}_Q^2  \tilde{q}_L^* +  \tilde{u}_R^T \tilde{m}_U^2  \tilde{u}_R^* + \tilde{d}_R^T \tilde{m}_D^2 \tilde{d}_R^*  + \tilde{l}_L^T \tilde{m}_L^2 \tilde{l}_L^*  \nonumber \\
 & + \tilde{e}_R^T \tilde{m}_E^2 \tilde{e}_R^* +  \tilde{m}_{H_u}^2 |H_u|^2 +  \tilde{m}_{H_d}^2 |H_d|^2 +  \tilde{m}_S^2 |S|^2 \nonumber \\
&  + A_U \tilde{q}_L^T y_U \tilde{u}_R^* H_u +  A_D \tilde{q}_L^T y_D \tilde{d}_R^* H_d + A_E \tilde{l}_L^T y_E \tilde{e}_R^* H_d   \nonumber \\
 & +  \lambda A_{\lambda} S H_u H_d + \kappa \frac{A_\kappa}{3} S^3 \, .
\end{align}
Without introducing additional couplings, the A-terms vanish at the messenger scale and gaugino masses and sfermion masses are given by
the usual minimal gauge mediation expressions:
\begin{align}
 M_i & =  N g_i^2 \tilde{m} \, ,  &
  \tilde{m}^2_{f}  & =  2N \sum_{i=1}^3 C_i(f)~g_i^4  \tilde{m}^2 \, ,
  \label{softMGM} 
\end{align}
where $\tilde{m} \equiv 1/(16 \pi^2) F/M$ and $C_i(f)$  is the quadratic Casimir of the representation of the field $f$ under ${\rm SU(3)}\times {\rm SU(2)}\times {\rm U(1)}$, for completeness given in Table 1. 
\begin{table}[ht]
\centering
\begin{tabular}{|c|ccccccc|}
\hline
& $Q$ & $U$  & $D$  & $L$  & $E$ & $H_u$  & $H_d$ \\
\hline
SU(3) & 4/3 & 4/3 & 4/3 & 0 & 0 & 0 & 0 \\
SU(2) & 3/4 & 0 & 0 & 3/4 & 0 & 3/4 & 3/4 \\
U(1) & 1/60 & 4/15 & 1/15 & 3/20 & 3/5 & 3/20 & 3/20 \\
\hline
\end{tabular}
\caption{Quadratic Casimirs of MSSM fields. }
\end{table}
In particular the soft mass of the singlet vanishes at the messenger scale, $\tilde{m}^2_{S}=0$. Together with the vanishing A-terms, this makes it difficult to trigger EWSB in minimal gauge mediation and motivates the introduction of additional
interactions among messengers and NMSSM fields.
\section*{1) DGS Model}
The following singlet-messenger couplings are added to the superpotential of Eq.~(\ref{NMSSM}):
\begin{align}
W_{\rm DGS} = S \left( \xi_D \Phi_{u}^{(1)} \Phi_{d}^{(2)}  +  \xi_T \Phi_{T}^{(1)} \Phi_{\overline{T}}^{(2)}   \right)  \, .
\label{DGS}
\end{align}
Notice that two copies of messengers are introduced (i.e.~$N=2$) in order to avoid that $S$ has the same quantum numbers of $X$, which would lead to tadpoles terms that destabilize the hierarchy. 
The superpotential in Eq.~(\ref{NMSSM}) and Eq.~(\ref{DGS}) is  the most general one  allowed by a $U(1)_Z \times Z_3$ symmetry with quantum numbers as in Table~\ref{DGScharges}.
\begin{table}[h!]
\centering
\begin{tabular}{|c|cccccc|}
\hline
& $X$ & $H_u$ &  $H_d$ & $S$   & $\Phi_{\bf 5}^1, \Phi_{\bf \overline{5}}^2$ & $ \Phi_{\bf \overline{5}}^1, \Phi_{\bf 5}^2$   \\
\hline
$U(1)_Z$ &  $1$ & $-1$ & $1$ & $0$ & $0$ & $-1$\\
\hline
$Z_3$ &  $0$ & $0$ & $2$ & $1$ & $1$ & $2$ \\
\hline
\end{tabular}
\caption{Charge assignments in the DGS model. \label{DGScharges}}
\end{table}
Note that one can impose a unification condition for $\xi_D$ and $\xi_T$ that allows to eliminate one parameter, 
\begin{align}
\xi_D(M_{\rm GUT}) = \xi_T(M_{\rm GUT}) = \xi \, .
\end{align}
The new couplings leads to additional contributions to soft terms, on top of the soft terms from minimal GM in Eq.~(\ref{softMGM}).  Now A-terms for the singlet are generated at one loop:
\begin{align}
A_{\lambda} &  =  -   \left( 2 \xi_D^2 + 3\xi_T^2 \right) \tilde{m}\, , \nonumber \\
A_{\kappa}  & =  - 3  \left( 2 \xi_D^2 + 3 \xi_T^2 \right)  \tilde{m} \, . 
\label{ADGS}
\end{align}
A soft mass for the singlet arises at one loop with an additional $F/M^2$ suppression that renders this contribution relevant only for very low messenger scales:
\begin{align}
 \tilde{m}_{S}^2|_{\rm 1-loop} = - 16 \pi^2 \tilde{m}^2 \frac{F^2}{M^4}  \frac{h(F/M^2)}{3} \left(2  \xi_D^2 + 3  \xi_T^2 \right), 
 \label{m1DGS}
 \end{align}
with the loop function
 \begin{equation}
 \label{hx}
h(x) \equiv \frac{3}{x^3} \log \frac{1-x}{1+x} - \frac{6}{x^4} \log (1-x^2)   = 1 + \frac{4}{5} x^2 + {\cal O}(x^4) \, .
\end{equation}
Unsuppressed soft masses in the Higgs sector are generated at two loops:
\begin{align}
\tilde{m}_{S}^2 & =   - \tilde{m}^2 \left[ \xi_D^2   \left(6/5 g_1^2 + 6 g_2^2\right) + \xi_T^2  \left(4/5 g_1^2 + 16 g_3^2\right) \right]   \nonumber \\
& - \tilde{m}^2 \left[ 4 \kappa^2 \left(2 \xi_D^2  + 3 \xi_T^2  \right) \right] \nonumber \\
& + \tilde{m}^2 \left[ 8 \xi_D^4  + 15 \xi_T^4  + 12 \xi_D^2 \xi_T^2  \right] \, , \nonumber \\
\Delta \tilde{m}_{H_u}^2 & = \Delta \tilde{m}_{H_d}^2 = - \tilde{m}^2 \lambda^2 \left( 2 \xi_D^2  + 3\xi_T^2  \right) \, .
\label{mDGS}
\end{align}

\section*{2) U Model}
In this model, we take just one messenger copy, $N=1$, and introduce the following couplings with $\Phi_{u}$:
\begin{align}
W_{\rm U} & = \lu  Q_3 U_3 \Phi_{u} +  \lsd S  \Phi_{u} H_d \, .
\label{WU}
\end{align}
The superpotential in Eq.~(\ref{NMSSM}) and Eq.~(\ref{WU}) is the most general one allowed by a $U(1)_Z \times Z_3$ symmetry with quantum numbers given in Table~\ref{Ucharges}, where $H_u$ is defined as that field that does not couple to $X$. 
\begin{table}[ht]
\centering
\begin{tabular}{|c|cccccc|}
\hline
& $X$ & $H_u$ &  $H_d$ & $S$   & $\Phi_{\bf 5}$ & $ \Phi_{\bf \overline{5}}$   \\
\hline
$U(1)_Z$ &  $1$ & $0$ & $0$ & $0$ & $0$ & $-1$\\
\hline
$Z_3$ &  $0$ & $1$ & $1$ & $1$ & $1$ & $2$ \\
\hline
\end{tabular}
\caption{Charge assignments in the U model. \label{Ucharges}}
\end{table}
One can also impose the condition 
\begin{align}
\lambda_{S_d} y_t = \lambda_t \lambda \, , 
\end{align}
 that would result from a explicit messenger-Higgs mixing~\cite{Evans:2011bea}. 

These couplings leads to additional contributions to soft terms, on top of the soft terms from minimal GM in Eq.~(\ref{softMGM}).  For one-loop A-terms one finds:
\begin{align}
(A_U)_{33} & = -3 \lu^2 \tilde{m} \, , \nonumber \\
(A_D)_{33} & = -   \left(   \lu^2  + \lsd^2\right) \tilde{m}\, , \nonumber \\
(A_E)_{33} & = -    \lsd^2 \tilde{m} \, , \nonumber \\
A_{\lambda} & =   - 3    \lsd^2 \tilde{m} \, ,\nonumber \\
A_{\kappa} & = - 6  \lsd^2 \tilde{m} \, .
\label{AU}
\end{align}
Also masses for $Q_3, U_3, S$ and $H_d$ are generated at one loop, but with an additional suppression by $F/M^2$:
\begin{align}
\Delta {\tilde m}^2_{Q_3}|_{\rm 1-loop}& =  - 16 \pi^2 \tilde{m}^2  \frac{F^2}{M^4}  \frac{h(F/M^2)}{6} \lambda_{t}^2 \, , \nonumber \\ 
\Delta {\tilde m}^2_{U_3}|_{\rm 1-loop} & =  - 16 \pi^2 \tilde{m}^2  \frac{F^2}{M^4}  \frac{h(F/M^2)}{3} \lambda_{t}^2 \, , \nonumber \\
 {\tilde m}^2_{S}|_{\rm 1-loop}& = - 16 \pi^2 \tilde{m}^2  \frac{F^2}{M^4}   \frac{h(F/M^2)}{3} \lambda_{S_d}^2 \, , \nonumber \\
\Delta {\tilde m}^2_{H_d}|_{\rm 1-loop}& = - 16 \pi^2 \tilde{m}^2  \frac{F^2}{M^4}   \frac{h(F/M^2)}{6} \lambda_{S_d}^2 \, \, , 
\label{m1U}
\end{align}
where the loop function $h(x)$ is defined in Eq.~(\ref{hx}). The two-loop soft masses are:
\begin{align}
\Delta {m}^2_{H_u} & =  - 3 \left(  3 y_t^2 \lu^2 +  \lsd^2 \lambda^2  \right) \tilde{m}^2 \, ,   \nonumber \\
\Delta {m}^2_{H_d} & =   6  \lambda \lsd \lu \yt  \tilde{m}^2   - \lsd^2  \left( 3/5 g_1^2 + 3 g_2^2 \right) \tilde{m}^2 \nonumber \\
& - 3 y_b^2 \lu^2 \tilde{m}^2 +  \lsd^2  \left(  2 \kappa^2  + 4 \lsd^2 + 2 \lambda^2 + 3 \lu^2  \right) \tilde{m}^2 \, , \nonumber \\
 \tilde{m}^2_S & =  \lsd^2  \left( - 6/5 g_1^2 - 6 g_2^2 - 8 \kappa^2 + 4 \lambda^2 + 8 \lsd^2  \right)  \tilde{m}^2 \nonumber \\
 & +  \left[  \lsd^2 \left( 6 \lambda_t^2+ 6 \yb^2+ 2 \ytau^2 \right) + 12 \lambda \lsd  \lu \yt  \right] \tilde{m}^2 \, ,  \nonumber \\
\Delta {\tilde m}^2_{Q_3} & =  \lu^2 \left(- 13/15 g_1^2 - 3 g_2^2  -16/3 g_3^2 + 6 \yt^2  \right) \tilde{m}^2 \nonumber \\
&  + \left(  6 \lu^4  + \lsd^2 \left(  \lambda_t^2- \yb^2 \right)  + 2 \lambda \lsd \lu \yt \right) \tilde{m}^2 \, , \nonumber \\ 
\Delta {\tilde m}^2_{U_3} & =\lu^2 \left(- 26/15 g_1^2 - 6 g_2^2 - 32/3 g_3^2 + 12 \yt^2  \right) \tilde{m}^2 \nonumber \\
& + \left( 12 \lu^4 +  2 \yb^2 \lu^2 +  2 \lsd^2  \lu^2 + 4 \lambda \lsd \yt \lu \right) \tilde{m}^2 \, , \nonumber \\
\Delta {\tilde m}^2_{D_3} & = - 2 \left( \yb^2 \lu^2  +  \lsd^2 \yb^2 \right) \tilde{m}^2\, , \nonumber \\
\Delta {\tilde m}^2_{L_3} & =  -  \lsd^2  \ytau^2 \tilde{m}^2 \, ,  \nonumber \\   
\Delta {\tilde m}^2_{E_3} & = - 2 \lsd^2 \ytau^2 \tilde{m}^2 \, .
\label{m2U}
\end{align}

\section*{3) DGSU Model}
In this model, we take two messenger copies, $N=2$, and introduce the following couplings:
\begin{align}
W_{\rm DGSU} & = S \left( \xi_D \Phi_{u}^{(1)} \Phi_{d}^{(2)}  +  \xi_T \Phi_{T}^{(1)} \Phi_{\overline{T}}^{(2)}   \right) \nonumber \\
&   + \lu  Q_3 U_3 \Phi_{u}^{(2)} +  \lsd S  \Phi_{u}^{(2)} H_d \, .
\label{WU}
\end{align}
The superpotential in Eq.~(\ref{NMSSM}) and Eq.~(\ref{WU}) is the the most general one allowed by a $U(1)_Z \times Z_3$ symmetry with quantum numbers given in Table~\ref{DGSUcharges}, where $H_u$ is defined as that field that does not couple to $X$. 
\begin{table}[ht]
\centering
\begin{tabular}{|c|cccccc|}
\hline
& $X$ & $H_u$ &  $H_d$ & $S$   & $\Phi_{\bf 5}^1,\Phi_{\bf \overline{5}}^2$ & $ \Phi_{\bf \overline{5}}^1,\Phi_{\bf 5}^2$   \\
\hline
$U(1)_Z$ &  $1$ & $-1$ & $1$ & $0$ & $0$ & $-1$ \\
\hline
$Z_3$ &  $0$ & $2$ & $0$ & $1$  & $1$ & $2$ \\
\hline
\end{tabular}
\caption{Charge assignment in the DGSU model. \label{DGSUcharges}}
\end{table}
In order to get the minimal number of parameters, we impose both the DGS unification condition at the GUT scale
\begin{align}
\xi_D(M_{\rm GUT}) = \xi_T(M_{\rm GUT}) = \xi \, .
\end{align}
and the Higgs-messenger mixing condition (for simplicity at the messenger scale)
\begin{align}
\lambda_{S_d} (M) y_t (M) = \lambda_t (M) \lambda (M) \, .
\end{align}
The superpotential couplings leads to additional contributions to soft terms, on top of the soft terms from minimal GM in Eq.~(\ref{softMGM}), from the DGS Model in Eq.~(\ref{ADGS}),(\ref{m1DGS}), (\ref{mDGS}) and the U model in Eq.~(\ref{AU}),(\ref{m1U}), (\ref{m2U}). These new terms are the mixed contributions of the form 
\begin{align}
\Delta {m}^2_{H_d} & = 3 \xi_T^2 \lsd^2 + 3 \xi_D^2 \lsd^2 \, , \nonumber \\
\Delta \tilde{m}^2_S & =  12 \xi_T^2 \lsd^2 + 16 \xi_D^2 \lsd^2 + 6  \xi_D^2 \lu^2 \, ,\nonumber \\
\Delta {\tilde m}^2_{Q_3} & = \xi_D^2 \lambda_t^2 \, , \nonumber \\
\Delta {\tilde m}^2_{U_3} & =  2 \xi_D^2 \lambda_t^2 \, .
\end{align}

\bibliographystyle{spphys}

\bibliography{bibliography}

\end{document}